\pdfoutput=1

\documentclass[pre,amsmath,amssymb,twocolumn,superscriptaddress]{revtex4}

\usepackage{graphicx}
\usepackage{bm}
\usepackage{braket}
\usepackage{afterpage}

\usepackage{subfigure}
\usepackage{chemfig}
\usepackage{slashed}

\newcommand{\beq}{\begin{equation}}
\newcommand{\eeq}{\end{equation}}

\newcommand{\aavg}[1]{\left\langle\!\left\langle #1 \right\rangle\!\right\rangle}

\newcommand{\ovl}[1]{\overline{#1}}

\newcommand{\xm}{x_{\min}}
\newcommand{\lb}{\lambda^\star}
\newcommand{\nsel}{\mathrm{ns}}
\newcommand{\sel}{\mathrm{s}}

\newcommand{\wf}{\omega_{\mathrm{ns}}}
\newcommand{\wu}{\omega_{\mathrm{s}}}
\newcommand{\pf}{p_{\mathrm{ns}}}
\newcommand{\pu}{p_{\mathrm{s}}}

\begin{document}

\title{Exploration-exploitation tradeoffs dictate the optimal distributions of phenotypes for populations subject to fitness fluctuations}

\author{Andrea De Martino}
\affiliation{Soft and Living Matter Laboratory, CNR-NANOTEC, 00185 Rome, Italy}
\affiliation{Italian Institute for Genomic Medicine, 10126 Turin, Italy}

\author{Thomas Gueudr\'e}
\affiliation{DISAT, Politecnico di Torino, 10129 Turin, Italy}

\author{Mattia Miotto}
\affiliation{Department of Physics, Sapienza University, 00185  Rome, Italy}

\begin{abstract}
We study a minimal model for the growth of a phenotypically heterogeneous population of cells subject to a fluctuating environment in which they can replicate (by exploiting available resources) and modify their phenotype within a given landscape (thereby exploring novel configurations). The model displays an exploration-exploitation trade-off whose specifics depend  on the statistics of the environment. Most notably, the  phenotypic distribution corresponding to maximum population fitness (i.e. growth rate) requires a non-zero exploration rate when the magnitude of environmental fluctuations changes randomly over time, while a purely exploitative strategy turns out to be optimal in two-state environments, independently of the statistics of switching times. We obtain analytical insight into the limiting cases of very fast and very slow exploration rates by directly linking population growth to the features of the environment. 
\end{abstract}

\maketitle

\section{\label{sec:Introduction}Introduction}

The exploration-exploitation trade-off scenario constitutes a paradigm for the optimal balance between the risky search for new resources and the safe exploitation of available ones that occurs in a variety of systems \cite{review}. As a generic example, one may consider a population occupying a patch of space in a land in which the availability of an essential resource fluctuates in time and across patches. By remaining on a certain patch for a sufficiently long time the population will be able to exploit the resource available in that patch to the fullest. That benefit, however, has to be weighed against the cost of the opportunities that are missed by not searching for a better patch. The central question concerns which balance of exploitation (stay) and exploration (go) will provide the population with the highest fitness (e.g. the fastest growth rate) in the long run. The optimal strategy is obviously interlocked with details like the statistics of resources and can be challenging to analyze at a quantitative level \cite{gdb,gm,dbm}. Still, fitness maximization is very often found to require a non-zero exploration rate. 

An especially significant effort to understand this trade-off is ongoing for biological systems, as seen e.g. in the recent interest about the ``ecology of cancer growth'' \cite{gore,swanton} (the strikingly diverse distributions of cell strains observed throughout different types of cancers) and its relationship to the timing of drug administration \cite{swanton}. Microbial systems have also been a natural testing ground for the exploration-exploitation scenario for many years. It is empirically known that, in fluctuating environments, microbes tend to display a high degree of phenotypic heterogeneity driven by stochasticity in the regulation of gene expression and metabolism \cite{hon,hallet,vdwb,tvo,rvso,vhwb}. The ability to explore the space of allowed phenotypes ultimately provides an effective route to hedge against environmental noise \cite{amvo,gssk}, favoring e.g. the persistence of a sub-population of resistant but slow-growing bacteria within a population subject to high doses of antibiotics \cite{bmck,kkl}. Starting with \cite{levins}, several mathematical models have shown that switching between different phenotypes at the individual cell level can be advantageous in rapidly changing conditions, depending essentially on (i) the statistics of environmental fluctuations and (ii) the specific coupling between the environment and the allowed phenotypes \cite{kl,gmr,vame,mhp,cgjd,mv,pk,sk,mmrw,hlg,wfm}. Such models capture the physical and mathematical complexity of these systems starting from minimal assumptions about the environment and/or the space of feasible phenotypes. In more structured cases, the spectrum of viable behaviors appears to be even richer \cite{gd}.

Here, inspired by recent work on single-cell physiology \cite{sattar} and by the growth-entropy balance that appears to underlie part of the empirical observations \cite{dcd}, we characterize the exploration-exploitation trade-off in a model for the growth of a phenotypically heterogeneous population in a fluctuating environment. In short, we assume that each phenotype is represented by an intrinsic or constitutive growth rate and that the landscape of phenotypes accessible to cells is described by a given probability distribution. Over time, cells modify their phenotype due, e.g., to stochastic fluctuations in intracellular composition or regulatory processes that effectively cause cells to perform random walks in the phenotypic landscape (the exploration part). In turn, the cellular replication rate is determined by the coupling to an externally varying environment. While fast phenotypes are in principle favored (the exploitation part), the environment is subject to fluctuations that can punish them (as e.g. in \cite{bmck}). In such conditions, the balance between exploration of the phenotypic space and exploitation of fast phenotypes ultimately controls both the overall fitness of the population and its structure (i.e. how individuals distribute over accessible phenotypes). 

We show that the optimal evolutionary strategy (yielding maximum fitness for the population) can indeed require a non-zero exploration rate as suggested by the general explore-exploit paradigm. The gain due to exploration is particularly marked in the most unpredictable environments. On the other hand, in  presence of more regular scenarios (e.g. periodic changes), an optimal population will adopt simpler strategies, such as maintaining two phenotypically distinct populations. 

Our analysis will focus on universal observables, relying on numerics for the general case. The limiting cases of very fast and very slow search rates will instead be characterized by approximate analytical arguments.

\section{Model definitions}

We consider a population of cells evolving in time. The phenotype of each cell is assumed to be fully characterized by a single variable $\lambda$, which we call the `constitutive replication rate' (CRR), taking on values in $[0,\lambda_{\max}]$. For sakes of simplicity, different values of $\lambda$ will effectively correspond to different cellular phenotypes. To account for the fact that some phenotypes might be easier to attain than others, the space of allowed phenotypes is assumed to be described by a probability density $q(\lambda)$ such that $q(\lambda)d\lambda$ represents the fraction of phenotypes with CRR between $\lambda$ and $\lambda+d\lambda$. The density of cells having CRR in $[\lambda,\lambda+d\lambda]$ at time $t$ is instead denoted by $n(\lambda,t)$. In turn, $N(t)=\int n(\lambda,t)d\lambda$ represents the total number of cells in the population at time $t$. Following e.g. \cite{dcd}, we assume that $n$ changes due to (a) replication events and (b) diffusion in the phenotypic space, whereby cells change their CRR from $\lambda$ to $\lambda'$. If the rate of the latter process is given by $W(\lambda\to\lambda')$, $n(\lambda,t)$ evolves according to
\begin{widetext}
\beq
\label{eq:master}
\frac{dn(\lambda,t)}{dt} = f(\lambda,t)\,n(\lambda,t) + \int\left[W(\lambda' \rightarrow \lambda)n(\lambda',t) - W(\lambda \rightarrow \lambda')n(\lambda,t)\right]d\lambda'~~,
\eeq 
\end{widetext}
where $f$ denotes the instantaneous replication rate (IRR) of cells with CRR $\lambda$. To couple the system to an external environment, we assume that the IRR depends both on the CRR $\lambda$ and on the state of an exogenously varying medium which, for sakes of simplicity, will be described by the single time-dependent variable $x$. To focus on a relevant case, we consider a fluctuating environment in which $x$ describes, in rough terms, the threshold fitness for replication under randomly occurring shocks. This corresponds to the choice
\beq
\label{eq:env}
f(\lambda,t) = \begin{cases}
\lambda &  \mbox{if} \quad \lambda \leq x(t) \\
0 & \mbox{otherwise} 
\end{cases}~~,
\eeq
according to which cells with CRR smaller than $x(t)$ can replicate at time $t$, while replication is inhibited for the others. 

To study the impact of randomness in the environment, we look at various scenarios, ranging from the most predictable (switching periodically between two fixed states) to the most random (switching after a random time and to a random value). More specifically, the threshold $x$ will fluctuate in time by switching between the value $x=\lambda_{\max}$, in which case all cells in the population can replicate, and a value $x =\lb<\lambda_{\max}$, in which case replication can only take place for cells with $\lambda \leq \lb$. We consider two choices for $\lb$. In the first case, $\lb$ is a constant kept fixed throughout the dynamics, so that $x$ takes the values $\lambda^\star$ and $\lambda_{\max}$ alternately, leading to a two-state environment (`const-$x$' case). In the second case, $\lb$ is sampled independently at every switch from a uniform distribution on the interval $[x_{\min}, \lambda_{\max}]$, leading to an environment with a continuum of states (`rand-$x$' case). For simplicity, we set $\lb=x_{\min}$ in the const-$x$ environment. Switches from the non-selective environment where all cells replicate to the selective one where only some do ($\nsel\to\sel$) and viceversa ($\sel\to\nsel$) are assumed to occur either periodically, i.e. after  fixed times $\wf$ and $\wu$ respectively (`const-$t$' case) or at exponentially distributed random times with means equal to $\wf$ or $\wu$ respectively (`rand-$t$' case). (We however expect all our results to be qualitatively robust to changes in the distributions from which times and thresholds are drawn.)

\begin{figure*}
\centering
\includegraphics[width= 0.95\textwidth]{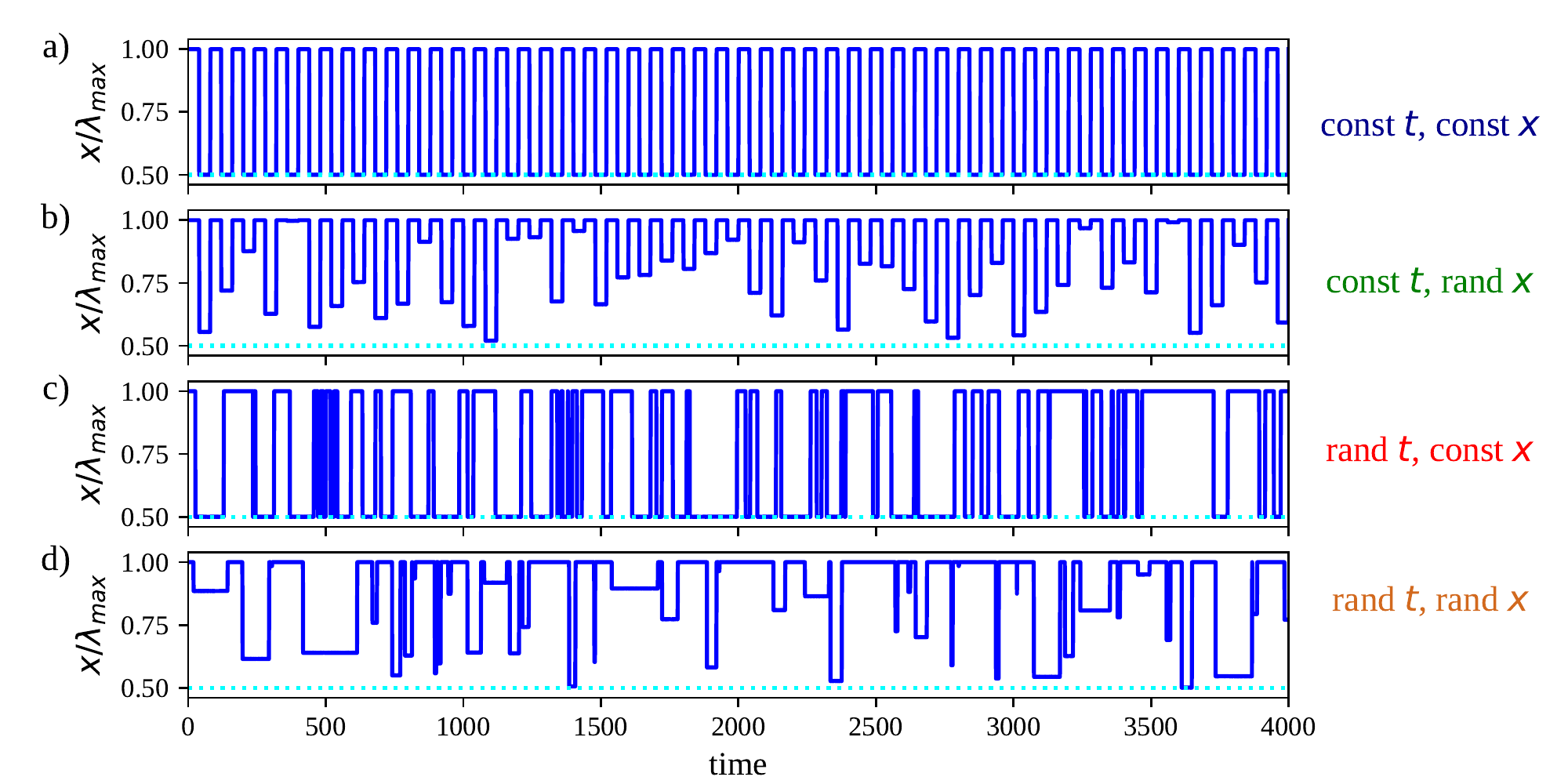}
\caption{
Representative behavior of the threshold $x$ as a function of time  (see Eq.~\ref{eq:env}) in the four environments we consider, namely \textbf{(a)} a periodic two-state environment where $x$ switches (in this case) between the values $\lambda_{\max}$ and $\xm=\lambda_{\max}/2$; \textbf{(b)} a periodically switching environment where $x$ takes on random values drawn uniformly from $[\xm,\lambda_{\max}]$; \textbf{(c)} a two-state environment where switches occur at exponentially distributed random times; \textbf{(d)} an environment where $x$ behaves as in (b) but in which switches occur at exponentially distributed random times. In this example, the characteristic switching times $\wf$ and $\wu$ are taken to be equal and fixed to 40 (a.u.). }
\label{fig:env}
\end{figure*}

Ultimately, for the process $x(t)$ we shall consider all possible mixtures of the above recipes for the threshold $x$ and the switching times (i.e. const-$t$, const-$x$; rand-$t$, const-$x$; etc.). In what follows, we begin by analyzing the simpler case of symmetric environment with $\wf = \wu=\omega$, representative examples of which are sketched in Fig. \ref{fig:env}. The asymmetric case with $\wf \neq \wu$ will be dealt with in  Sec.~\ref{sec:uneven}. 

Introducing the population density
\beq
\label{eq:p_n}
p(\lambda,t) = \frac{n(\lambda,t)}{N(t)}~~,
\eeq 
we re-cast Eq. (\ref{eq:master}) as 
 \begin{widetext}
\beq
\label{eq:masterP}
\frac{dp(\lambda, t)}{dt} = \Big[ f(\lambda,t) - \mathbb{E}_{\lambda\leq\lambda_{\max}} f(\lambda,t)\Big] p(\lambda,t) + \int \Big[W(\lambda' \to \lambda)p(\lambda',t) - W(\lambda \to \lambda')p(\lambda,t)\Big]d\lambda'~~,
\eeq
 \end{widetext}
with
\begin{gather}
\label{eq:Nric}
\mathbb{E}_{\lambda\leq\lambda_{\max}} f(\lambda,t) = \int_0^{\lambda_{\max}} f(\lambda,t)p(\lambda,t) d\lambda ~~.
\end{gather}
Furthermore, we assume that transition rates satisfy a detailed-balance condition of the form 
\beq
\label{eq:DetB}
W(\lambda \rightarrow \lambda')q(\lambda) = W(\lambda' \rightarrow \lambda)q(\lambda')~~,
\eeq
with $q(\lambda)$ the density of phenotypes, and introduce the mean waiting time $\tau$ characterizing transitions via
\beq
\label{eq:tau}
\int  W(\lambda \rightarrow \lambda') d\lambda'= \frac{1}{\tau}~~.
\eeq
We further assume that only transitions from phenotype $\lambda$ to phenotypes $\lambda \pm \delta\lambda$ are allowed, with equal probability and small $\delta\lambda$ (`diffusive transition kernel'). This choice provides the most natural route to model the effects induced at phenotypic level by small random  fluctuations in intracellular composition, as they are unlikely to cause major gains or losses in terms of CRR. One easily shows (see Appendix~\ref{app:FP} and \cite{dcd}) that Eq. (\ref{eq:masterP}) in this case can be approximated with the non-linear Fokker-Planck equation
\begin{widetext}
\beq
\label{eq:Pdiff}
\frac{d p(\lambda,t)}{dt} = \Big[ f(\lambda,t) - \mathbb{E}_{\lambda\leq\lambda_{\max}} f(\lambda,t)\Big]p(\lambda,t) + D\left[ \frac{\partial^2 p(\lambda,t)}{\partial^2 \lambda} - \frac{\partial}{\partial \lambda} \left[ p(\lambda,t)\frac{\partial}{\partial\lambda}(\ln q(\lambda))\right]\right]~~,
\eeq
\end{widetext}
where  $D = \frac{(\delta\lambda)^2}{2\tau}$ is the diffusion coefficient in the phenotypic space. 

We finally have to specify a form for the phenotypic landscape $q(\lambda)$. To focus on a realistic case, we set
\beq
\label{eq:qlambda}
q(\lambda) = \frac{a+1}{\lambda_{\max}}\left(1-\frac{\lambda}{\lambda_{\max}}\right)^a ~~,
\eeq
where the exponent $a\geq 0$ modulates the steepness of $q(\lambda)$. In short, the larger $a$, the more heterogeneous the landscape, with slow phenotypes being increasingly more frequent than fast ones as $a$ increases.  The above choice is based on recent studies showing that functions like (\ref{eq:qlambda}) describe the CRR landscape underlying genome-scale models of bacterial metabolic networks, with values of $a$ extracted from genome-scale models of {\it E.coli} lying between 200 and 300 depending on the specifics of the  environment \cite{dcd,dcd2}. To focus on tractable extremes, we shall consider explicitly the cases $a=0$ (uniform $q(\lambda)$) and $a=20$ (strongly heterogeneous $q(\lambda)$).

The setup just described generalizes that considered in \cite{dcd,dm} to the case in which the instantaneous replication rate $f$ depends on the coupling of cells to a fluctuating environment. The structure of a population governed by (\ref{eq:Pdiff}) emerges from the balance between the term that rewards fast-growing states (which are however sensitive to environmental shocks) and the diffusion term favoring states with larger entropy in the phenotypic space (but slower replication rates). In the following, we characterize the above setting from the viewpoints of
\begin{enumerate}
\item how the interplay between replication and diffusion (i.e. the trade-off between exploration and exploitation) affects the growth rate of the population as a whole;
\item the emergent asymptotic structure of the population, i.e. how cells distribute over the one-dimensional phenotypic space $[0,\lambda_{\max}]$ at long times.
\end{enumerate}
It is important to note that, in symmetric environments, two different timescales rule the time evolution of $p(\lambda,t)$:   the mean switching time between different environments ($\omega$) and the mean time to transition between different phenotypes ($\tau$). The latter is inversely proportional to the diffusion constant $D$. The system's behaviour is ultimately modulated by the ratio $\omega/\tau$. To explore the full range of this ratio, it is convenient to fix one time scale, e.g. $\omega$, and use the other (i.e. $D$) as a control parameter. The limiting cases $\omega\ll \tau$ (in which exploration occurs on much longer time scales than exploitation) and $\omega\gg \tau$ (in which exploration occurs on much shorter time scales than exploitation) correspond to $D\to 0$ and $D\gg 1$, and we shall refer to these as the `exploitation' and `exploration' limits, respectively.

\section{Results}

\subsection{Dynamical patterns of population structure under symmetric switching}

\begin{figure*}[]
\centering
\includegraphics[width= 0.95\textwidth]{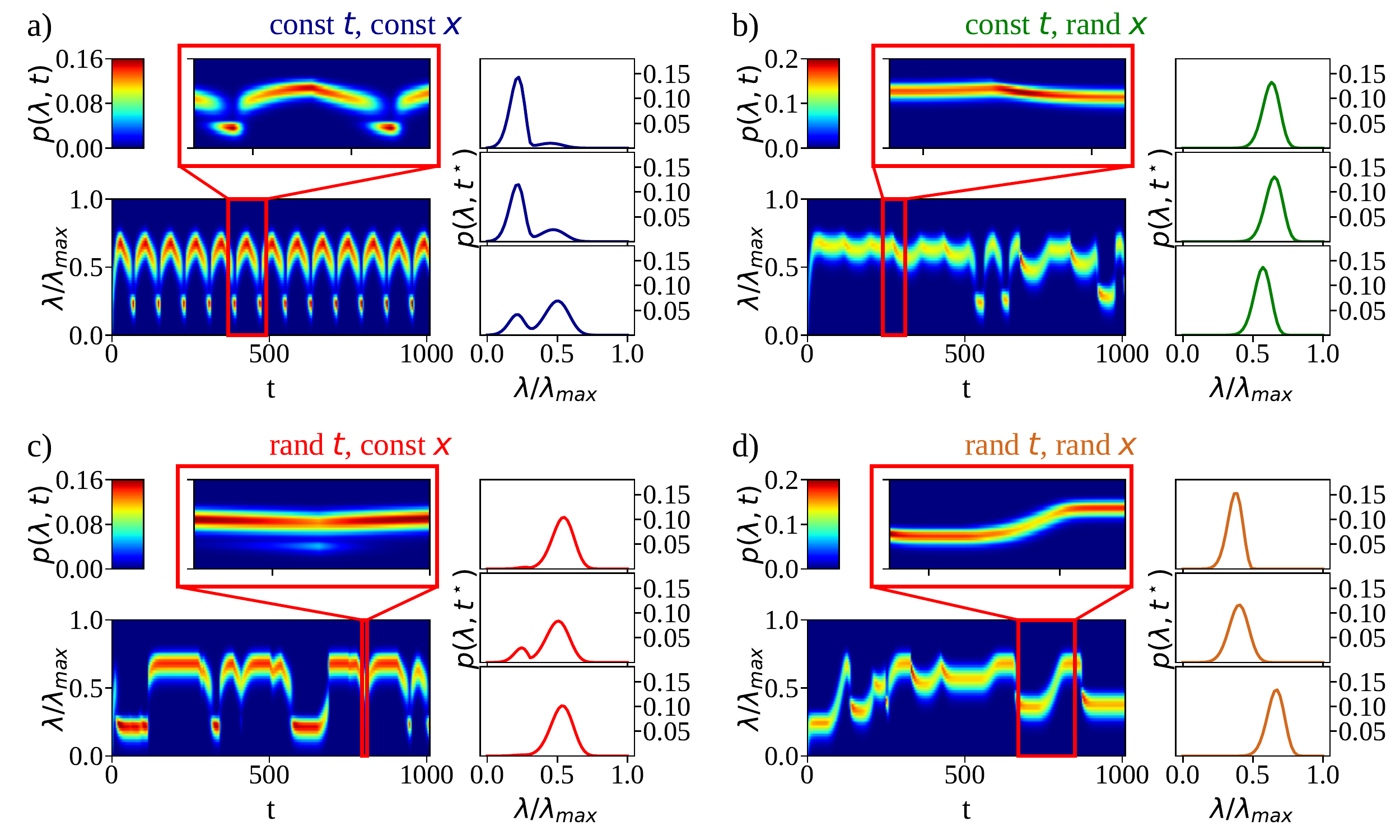}
\caption{Colormaps showing representative probability densities $p(\lambda,t)$ obtained by solving Eq.~(\ref{eq:Pdiff}) numerically in the different kinds of environment described in Fig. \ref{fig:env}. Panels to the right of each map depict the density profile at different time points within the zoomed-in region, at time increasing from top to bottom. Results are shown for \textbf{a)} const-$t$ and const-$x$ environment, \textbf{b)} const-$t$ and rand-$x$ environment, \textbf{c)} rand-$t$ and const-$x$ environment, and \textbf{d)} rand-$t$ and rand-$x$ environment. Parameter values: $a = 20$, $x = 0.3\lambda_{\max}$, $D = 10^{-3}$.} 
\label{fig:Nmaxima_vs_D}
\end{figure*}

The non-linear Fokker-Planck equation~(\ref{eq:Pdiff}) can be solved numerically for any choice of the environment, of the diffusion coefficient and of the prior phenotypic density $q(\lambda)$. After a short transient, $p(\lambda,t)$ appears to settle in qualitative robust, environment-dependent patterns, a sample of which is shown in Fig. \ref{fig:Nmaxima_vs_D}. 
Different types of distributions emerge across the various environments, including bimodal distributions in which most of the population occupies the two peaks alternately (panel a) or in which one peak always dominates over the other (panel c), unimodal distributions with fluctuating positions (panel b) and unimodal distributions in which peaks drift in a specific direction (panel d). While all of these can occur in every type of environment, both their frequency of occurrence and the relative intensities of the peaks appear to be strongly environment-dependent. 

Such patterns provide hints about the way in which the population copes with environmental fluctuations. An important feature observed from data is that, independently of whether switches occur periodically or randomly, adaptation to two-state environments (const-$x$) is achieved more efficiently by structuring the population in a bimodal form, while complex environments (rand-$x$) favor unimodal distributions. We shall see in the following that such a scenario is indeed correct even asymptotically, although it can be modulated by the strength of diffusion.

\subsection{Population growth rate and statistics at long times under symmetric switching}

\begin{figure*}
\includegraphics[width=0.95\textwidth]{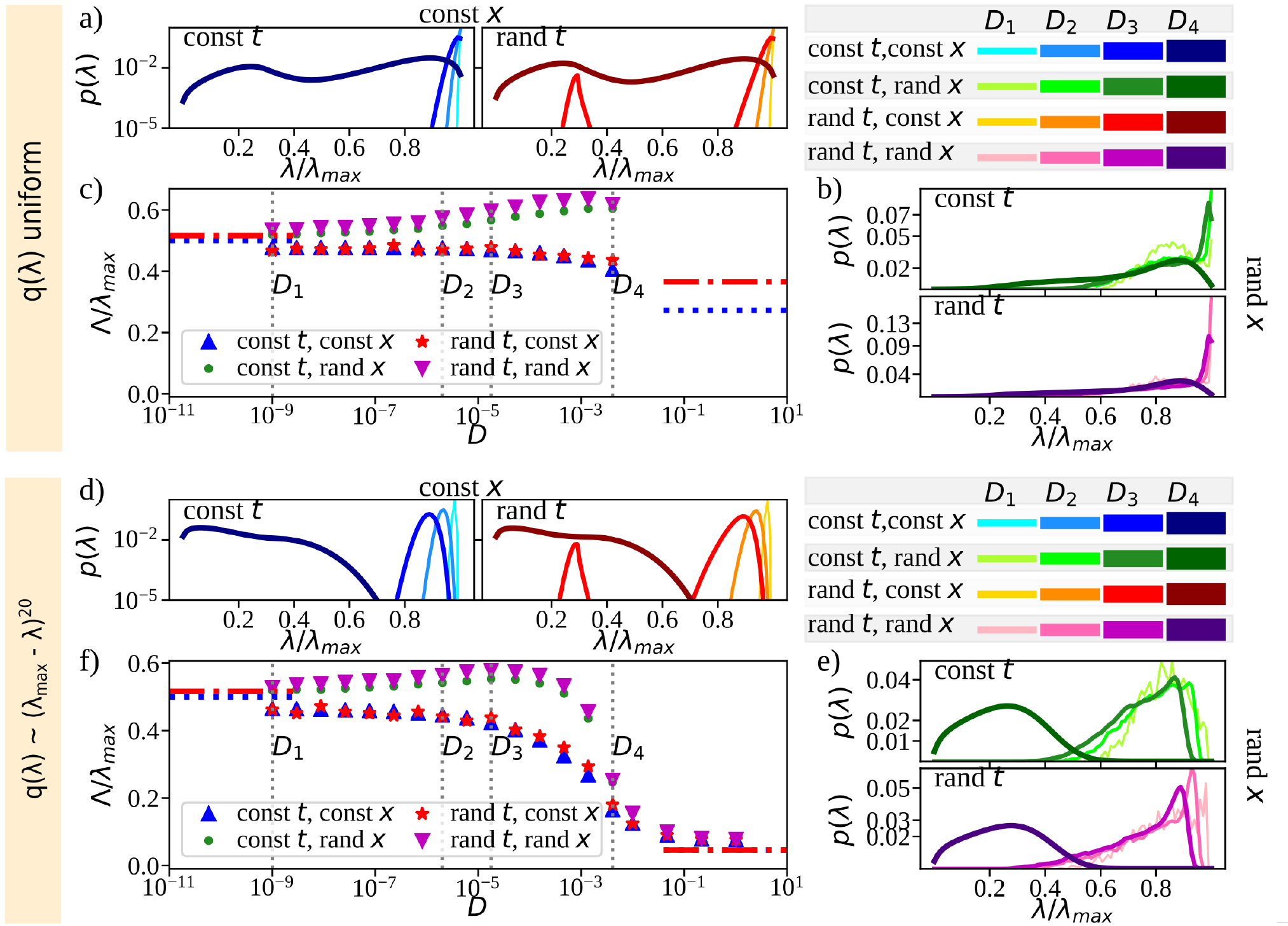}
\caption{\textbf{(a and b)} Asymptotic, time-averaged phenotypic distributions obtained for a population evolving according to Eq.~(\ref{eq:Pdiff}) with a diffusive kernel in a uniform background phenotypic landscape $q(\lambda)$ in const-$x$ (panel (a)) and rand-$x$ (panel (b)) environments for different time scenarios and values of $D$, marked by increasing color shades and line widths. \textbf{(c)} Asymptotic population growth rate $\Lambda$ (in units of $\lambda_{\max}$) as a function of $D$ for the four types of environment. Vertical dotted lines mark the values of $D$ studied in panels (a) and (b). Horizontal lines at small and large $D$ stand for the analytical estimates for $\Lambda$ obtained in the const-$x$ regime (dotted blue line, Eqs (\ref{dimeq}) for small $D$ and (\ref{eq:Dinf}) for large $D$) and the rand-$x$ regime (dot-dashed red line, Eqs (\ref{ventinove}) for small $D$ and (\ref{eq:Dinf}) for large $D$), respectively. \textbf{(d to f)} Same as a--c but with $q(\lambda)$ as in Eq. (\ref{eq:qlambda}) (with $a=20$) rather than uniform. Displayed curves are averaged over 100 independent realizations of the dynamics performed with $\xm = 0.3\lambda_{\max}$. \label{fig:d_a0_20_p03}}
\end{figure*}

As we are mostly interested in understanding how the system behaves in the long-time limit, we focus on the long-term population structure as well as on the growth rate
\begin{eqnarray}
\label{eq:Lam}
\Lambda &\equiv& \lim_{t\to\infty}\frac{1}{t} \ln\frac{N(t)}{N(0)}\\ 
& = &\lim_{t\to\infty}\frac{1}{t} \int_0^t \big[\mathbb{E}_{\lambda\leq\lambda_{\max}} f(\lambda,t')\big] dt'~~.
\end{eqnarray}
(The second equality follows directly from Eq. (\ref{eq:master}) and from the fact that $N(t)=\int n(\lambda,t)d\lambda$.) $\Lambda$ will be used as a proxy for the long-term evolutionary success of the population. Figure \ref{fig:d_a0_20_p03} shows, for all environments, the stationary probability distributions $p(\lambda)$ obtained by averaging over time after $\Lambda$ has reached its stationary value, for representative values of the parameters (in particular for $x_{\min}=0.3\,\lambda_{\max}$, describing a strong negative perturbation which can be evaded only by cells whose CRR is at most 30\% of the maximum), different values of $D$, and for $a=0$ (corresponding to a uniform phenotypic landscape, top panels) and $a=20$ (a strongly heterogeneous landscape with a predominance of slow growing states, bottom panels). 

Generically, at sufficiently small values of $D$, phenotypes tend to concentrate close to $\lambda_{\max}$ (see Fig. \ref{fig:d_a0_20_p03}a,b and d,e). This situation reproduces the `exploitation' limit $D\to 0$, where (\ref{eq:Pdiff}) reduces to the replicator dynamics
\begin{gather}
\label{eq:RepDyn}
\frac{dp(\lambda, t)}{dt} =  \Big[ f(\lambda,t) - \mathbb{E}_{\lambda\leq\lambda_{\max}} f(\lambda,t)\Big]p(\lambda,t)~~.
\end{gather}
A population whose phenotypic diffusion occurs on exceedingly long time scales (compared to those characterizing environmental fluctuations) can only grow exploiting resources available from the environment and is therefore maximally sensitive to environment-derived shocks.  In such a case, the population growth rate is significantly smaller than $\lambda_{\max}$, see Fig. \ref{fig:d_a0_20_p03}c and f, due to the growth-curbing effect of environmental fluctuations. (We shall analyze this limit at quantitative level in the following.) 

Upon increasing $D$ (and therefore the relevance of diffusion in the phenotypic space), distributions start to acquire non-trivial traits, including bimodality (see Fig. \ref{fig:d_a0_20_p03}a,d) and extended tails (see Fig. \ref{fig:d_a0_20_p03}b,e). The population growth rate $\Lambda$ then increases with $D$ with respect to the small-diffusion limit in complex (rand-$x$) environments, where the population structure develops tails. In such cases, $\Lambda$ has a well-defined maximum at a specific value of $D$ (which depends, as in \cite{gdb}, on the characteristic time of environmental switches), marking the existence of an optimal trade-off between diffusion (exploration) and growth (exploitation) in the given environment. On the other hand, the population growth rate decreases continuously with $D$, albeit slowly, in the simpler two-state (const-$x$) environments, implying that any amount of exploration is detrimental to fitness in such contexts.

When diffusion dominates the dynamics (larger values of $D$), $\Lambda$ appears to drop rapidly in all environments. In such a  case, which is close to the purely `exploration' limit $D\to\infty$ that is analyzed in detail below, cells explore the phenotypic space very efficiently, continuously redistributing their CRR among allowed states. The asymptotic behavior is hence dominated by the background provided by $q(\lambda)$. Indeed, the phenotypic distribution evolves towards its stationary limit $q(\lambda)$ due to the detailed balance constraint (\ref{eq:DetB}).

These results suggest that phenotypic diffusion can indeed be tuned to cope optimally with environmental fluctuations so as to ensure a significant gain in terms of fitness, provided the selective threshold of the environment changes randomly over time. In such a case, the fitness advantage appears to be slightly more marked when $a$ is smaller. Still, the qualitative scenario just described is  robust to changes in $a$. Correspondingly, the population structures into an extended unimodal distribution of phenotypes. On the other hand, in an environment fluctuating between two well-defined states, bimodal phenotypic distributions occur but exploration does not appear to provide a significant fitness advantage. 

Note that a similar qualitative scenario for $\Lambda$ is obtained for weaker environmental perturbations (i.e. larger $x_{\min}$), the main effect induced by increasing $x_{\min}$ being (expectedly) that of reducing the gap in $\Lambda$ as a function of $D$ between const-$x$ and rand-$x$ environments without modifying the overall behaviour of individual cases.

\subsection{Exploitation limit (case of symmetric switching)}\label{ssec:envlim}

\begin{figure}[]
\centering
\includegraphics[ width= \columnwidth]{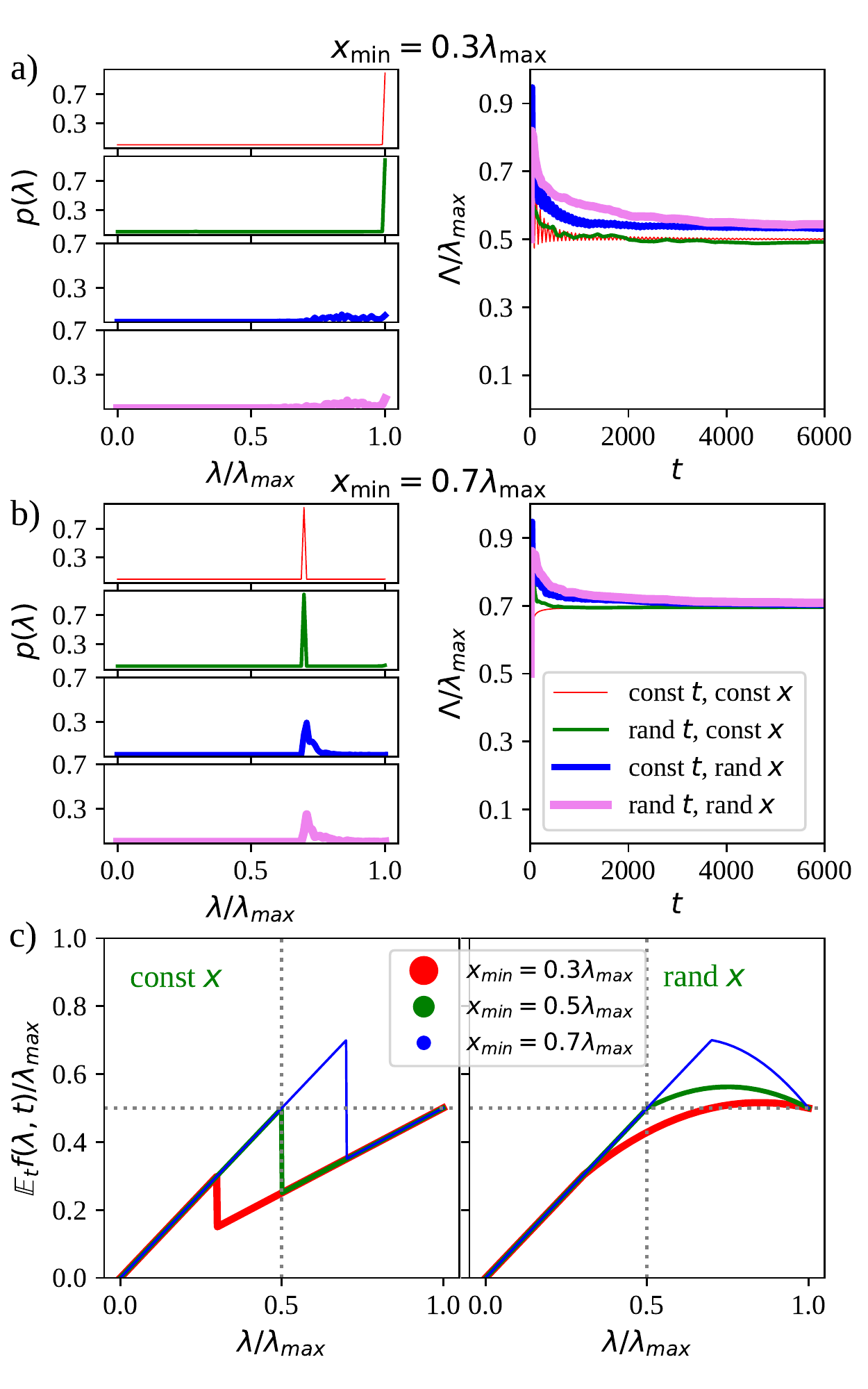}
\caption{\textbf{(a and b)} Long-time phenotypic distributions (left) and time evolution of the population growth rate $\Lambda$ (right) in the absence of diffusion in the different environments (represented by different colors and line widths) and for $x=0.3\lambda_{\max}$ (panel (a)) and $x=0.7\lambda_{\max}$ (panel (b)). In the former case ($x<\lambda_{\max}/2$), the distribution can achieve the highest possible CRR. In turn, the long term fitness $\Lambda$ sets around $\lambda_{\max}/2$. In the latter case ($x>\lambda_{\max}/2$), the distribution peaks around the threshold CRR, while the population achieves a growth rate $\Lambda$ larger than $\lambda_{\max}/2$. Curves are averaged over 100 realizations of the dynamics. \textbf{(c)} Time-averaged mean $\mathbb{E}_t f(\lambda,t)$ of $f$  as a function of the CRR for const-$x$ (left) and rand-$x$ (right) environments and for three different values of $x_{\min}$. One sees that the position of the maximum depends both on the chosen threshold and on the specific environment. } 
\label{fig:Repdata}
\end{figure}

To characterize our model in greater detail, it is convenient to focus on its limiting behaviors starting from the case $D\to 0$ (i.e. exploitation much faster than exploration), in which (\ref{eq:Pdiff}) reduces to (\ref{eq:RepDyn}). Here, the population is dominated by the cells carrying the largest CRR. Intuitively, though, the coupling to the environment limits the reproductive efficiency of fast-growing phenotypes and ultimately introduces cut-offs to the CRR that are represented in the population. The parameter controlling this effect is $x_{\min}$. Numerical results indeed show (see Fig.~\ref{fig:Repdata}) that, while the statistics of switching times does not appear to qualitatively influence the long-time limit, const-$x$ and rand-$x$ regimes produce qualitatively different asymptotics for the phenotypic distribution depending on whether $x_{\min}<\lambda_{\max}/2$ (panel a) or $x_{\min}>\lambda_{\max}/2$ (panel b). 
Specifically, for $\xm>\lambda_{\max}/2$ (weaker perturbation) the population concentrates around $\lambda=\xm$ in every environment with an overall fitness $\Lambda>\lambda_{\max}/2$. For $\xm<\lambda_{\max}/2$, instead, $p(\lambda)$ displays a peak at $\lambda=\lambda_{\max}$ in const-$x$ environments while an extended set of phenotypes is represented in the population when $x$ is random and uniform. In both cases, the population growth rate $\Lambda$ settles close to $\lambda_{\max}/2$. In other words, more efficient phenotypes appear to dominate  the population when the perturbation is stronger and the overall growth rate is slower, while a weaker perturbation leading to a larger population growth rate seems to select for less efficient phenotypes.  

A key observation to understand these results is that, independently of whether environmental switches occur at fixed times or at random times, in a symmetric environment with $\wf=\wu$ cells will spend on average half the time in the ``favorable'', not selective environment with $x=\lambda_{\max}$ and the other half in the selective environment with $x<\lambda_{\max}$. The statistics of switching times should therefore not be expected to influence outcomes at least as long as averages are concerned. On the other hand, because $x$ is a random variable, the IRR $f$ (Eq. \ref{eq:env}) will also be randomly fluctuating in time, with a mean value given by
\begin{equation}
\label{eq:bar_f}
\mathbb{E}_t f(\lambda,t) = \lim_{t\to\infty} \frac{1}{t}\int_0^t f(\lambda,t') dt'~~.
\end{equation}
It is now convenient to discuss the const-$x$ and rand-$x$ cases separately.

\subsubsection{Const-$x$ (two-state) environments}

For the `const-$x$' case (two-state environment with $x$ oscillating between $\lambda_{\max}$ and a constant value $\xm$), $f$ will equal $\lambda$ at all times if $\lambda<\xm$, while for $\lambda>\xm$ it will be equal to $\lambda$ for approximately half the time and to zero for the other half. This implies that
\begin{equation}
\label{eq:fc}
\mathbb{E}_t f(\lambda,t) \simeq 
\begin{cases}
\lambda & \mbox{if}~\lambda<\xm~~,\\
\lambda / 2 & \mbox{if}~\lambda\geq \xm~~.\\
\end{cases}\\
\end{equation}
The mean IRR therefore displays a discontinuity at the threshold $\xm$, and the value of $\lambda$ for which it attains a maximum depends on the value of $\xm$ (see Fig.~\ref{fig:Repdata}c, left panel). In specific, for $\xm>\lambda_{\max}/2$ (resp. $\xm < \lambda_{\max}/2$), the mean IRR has a maximum for $\lambda = \xm$ (resp. $\lambda = \lambda_{\max}$), where $\mathbb{E}_t f(\lambda,t)=\xm$ (resp. $\mathbb{E}_t f(\lambda,t)=\lambda_{\max}/2$). Hence, at long times, we expect the population to grow at the fastest IRR achievable, with a phenotypic distribution $p(\lambda)$ peaked at $\lambda = \xm$ (resp. $\lambda = \lambda_{\max}$) for $\xm>\lambda_{\max}/2$ (resp. $\xm<\lambda_{\max}/2$). This is in agreement with the numerical evidence shown in Fig.~\ref{fig:Repdata}a,b (as well as in Fig. \ref{fig:d_a0_20_p03}) for const-$x$ environments.

Based on the above reasoning we can approximate $p(\lambda,t)$ with the bimodal function
\beq
\label{eq:P_bi}
p(\lambda,t) \simeq \alpha(t)\delta(\lambda-\lb)+(1-\alpha(t))\delta(\lambda-\lambda_{\max})~~,
\eeq 
with $0\leq\alpha(t)\leq 1$ a time-dependent coefficient quantifying the fraction of cells with CRR equal to $\lambda^\star<\lambda_{\max}$. (For sakes of simplicity, we shall henceforth omit to indicate explicitly the dependence of $\alpha$ on time.) We can then use (\ref{eq:RepDyn}), which in discrete time takes the form
\begin{multline}
p(\lambda,t+\delta t)\simeq \\ \Big\{1+\Big[f(\lambda,t)-\mathbb{E}_{\lambda\leq\lambda_{\max}} f(\lambda,t)\Big]\delta t\Big\}\, p(\lambda,t)~~,
\end{multline}
to evolve the above ansatz for small time intervals $\delta t$ during which the environment does not change. 

In non-selective conditions ($x = \lambda_{\max}$), one can use the fact that 
\begin{equation}
\mathbb{E}_{\lambda\leq\lambda_{\max}} f(\lambda,t)= \alpha \lb + (1-\alpha)\lambda_{\max}
\end{equation}
to arrive at
\begin{eqnarray}
\label{eq:p_fav}
p(\lambda,t+\delta t) &\simeq&  
\left(\alpha - \delta\alpha_{\rm ns}\right)\,\delta(\lambda-\lb)  +\nonumber\\ & & +\left(1-\alpha + \delta\alpha_{\rm ns}\right)\,\delta(\lambda-\lambda_{\max})~~,
\end{eqnarray}
where $\delta\alpha_{\rm ns} = (\lambda_{\max} - \lb)(1-\alpha)\delta t$.  

In a selective environment ($x = \lb$), instead,  
\begin{equation}
\mathbb{E}_{\lambda\leq\lambda_{\max}} f(\lambda,t) = \alpha\lb~~,
\end{equation}
and one finds
\begin{multline}
\label{eq:P_bi_evo_stress}
p(\lambda,t+\delta t) \simeq  \left(\alpha + \delta\alpha_{\rm s}\right)\delta(\lambda-\lb)  +\\+ \left(1-\alpha - \delta\alpha_{\rm s}\right)\delta(\lambda-\lambda_{\max})~~,
\end{multline}
with $\delta\alpha_{\rm s} = \lb(1-\alpha)\delta t$. 

This shows that, at every switch, the population distribution will tend to shift from one threshold to the other, but the speed with which the two peaks grow or shrink are different. In particular, one has
\beq
\frac{\delta\alpha_{\rm ns}}{\delta\alpha_{\rm s}} = \frac{(\lambda_{\max} - \lb)}{\lb}~~.
\eeq 
This implies that $\delta\alpha_{\rm ns}<\delta\alpha_{\rm s}$ for $\lb>\lambda_{\max}/2$. Hence the peak growing at speed   $\delta\alpha_{\rm s}$ is favored and the probability density will peak around $\lb$ in the long run. On the other hand, $\delta\alpha_{\rm ns}>\delta\alpha_{\rm s}$ when $\lb<\lambda_{\max}/2$, causing the population to concentrate around $\lambda_{\max}$. In other terms, 
\beq
\label{eq:p_delta}
p(\lambda) \simeq
\begin{cases}
\delta(\lambda- \lb) & \rm if~~\lb > \lambda_{\max}/2\\
\delta(\lambda- \lambda_{\max}) & \rm if~~\lb < \lambda_{\max}/2
\end{cases}~~,
 \eeq
in agreement with the numerical picture for the two-state (const-$x$) environment shown in Fig. \ref{fig:Repdata}. 
 
This result can be used to obtain an analytical approximation for $\Lambda$. In fact, considering that the system spends roughly half the time in the non-selective environment ($x=\lambda_{\max}$) and the other half in the selective one ($x=x_{\min}$), we have (see (\ref{eq:Lam}))
\begin{eqnarray}\label{ventitre}
\Lambda &\simeq& \frac{1}{2}\, \mathbb{E}_{\lambda\leq\lambda_{\max}} f(\lambda,t) + \frac{1}{2} \,\mathbb{E}_{\lambda\leq x_{\min}} f(\lambda,t)\\ & \simeq & \begin{cases}
\lb & \mbox{if } \lb > \frac{\lambda_{\max}}{2}\\
\frac{\lambda_{\max}}{2} & \mbox{if } \lb \leq \frac{\lambda_{\max}}{2}
\end{cases}~~,\label{dimeq}
\end{eqnarray}
where 
\beq
\mathbb{E}_{\lambda\leq z} f(\lambda,t)=\int_0^z f(\lambda,t)p(\lambda,t)d\lambda 
\eeq
and we used the fact that $\mathbb{E}_{\lambda\leq\lambda_{\max}} f(\lambda,t)=\lb$ (resp. $\mathbb{E}_{\lambda\leq\lambda_{\max}} f(\lambda,t)=\lambda_{\max}$) for $\lb>\lambda_{\max}/2$ (resp. $\lb<\lambda_{\max}/2$), while $\mathbb{E}_{\lambda\leq x_{\min}} f(\lambda,t)=\lb$ (resp. $\mathbb{E}_{\lambda\leq x_{\min}} f(\lambda,t)=0$) for $\lb>\lambda_{\max}/2$ (resp. $\lb<\lambda_{\max}/2$).

In Fig.~\ref{fig:d_a0_20_p03}c, we show that the value of $\Lambda$ estimated numerically agrees with the one just derived in the limit $D\to 0$ (horizontal blue line) for $\lambda^\star=x_{\min}$. Note that $\Lambda$, Eq. (\ref{dimeq}), corresponds to the maximum of the time-averaged IRR $\mathbb{E}_t f(\lambda,t)$ (see Fig.~\ref{fig:Repdata}c), confirming how, for small $D$ (when diffusion is much slower than environmental fluctuations), fitness is ultimately limited by the environment alone.

\subsubsection{Rand-$x$ environments}

In the `rand-$x$' case ($x$ oscillating between $\lambda_{\max}$ and a random value $\lb$ uniformly chosen from $[x_{\min},\lambda_{\max}]$), $f$ will again equal $\lambda$   roughly half the time, while for the other half it will be randomly zero or $\lambda$ depending on $\xm$. In particular,  ${\rm Prob}\{f=\lambda\}\equiv{\rm Prob}\{x > \lambda\}=1-\phi$, with
\begin{equation} 
\phi=\frac{\lambda - x_{\min}}{\lambda_{\max}-x_{\min}}~~.
\end{equation}
One therefore finds
\begin{equation}
\label{eq:fu}
\mathbb{E}_t f(\lambda,t) \simeq 
\begin{cases}
\lambda & \mbox{if}~\lambda<\xm\\
\lambda\left( 1 - \frac{\phi}{2}\right) & \mbox{if}~\lambda\geq \xm
\end{cases}~~,
\end{equation}
from which one sees that $\mathbb{E}_t f(\lambda,t)$ attains a maximum value $\ovl{f}_{\max}$ given by 
\beq
\ovl{f}_{\max} = \frac{1}{2}\frac{\left(\lambda_{\max} - \frac{1}{2}\,x_{\min}\right)^2}{\lambda_{\max} - x_{\min}}~~,
\eeq
at $\lambda = \lambda_{\max} - \frac{1}{2}\,x_{\min}$ if  $x_{\min}<\frac{2}{3}\lambda_{\max}$, while $\ovl{f}_{\max}=x_{\min}$ at $\lambda = x_{\min}$ if $x_{\min}>\frac{2}{3}\lambda_{\max}$. 
In complete analogy with the previous case, the population concentrates around phenotypes $\lambda$ for which $\mathbb{E}_t f(\lambda,t)$ is maximum, while for the asymptotic growth rate of the population $\Lambda$ one finds
\beq\label{ventinove}
\Lambda\simeq \ovl{f}_{\max} ~~,
\eeq
(see Fig.~\ref{fig:Repdata}). The results displayed in Fig.~\ref{fig:d_a0_20_p03}c (red horizontal line for $D\to 0$) indeed support this conclusion.

\subsection{Exploration limit (case of symmetric switching)}\label{sssec:highLim}

In the limit  $D\rightarrow \infty$ (and more generally whenever  diffusion occurs on time scales much faster than those of environmental fluctuations), the growth term in Eq. (\ref{eq:masterP}) is negligible with respect to the diffusion one and population is rapidly redistributed according to the underlying phenotypic landscape described by $q(\lambda)$. As a consequence   $p(\lambda) \to q(\lambda)$ asymptotically. It is again possible to derive an approximate expression for $\Lambda$ from Eq.~(\ref{eq:Lam}) following the lines traced in the previous section. One finds, in analogy with (\ref{ventitre}), 
\beq\label{Lamd}
\Lambda \simeq \frac{1}{2}\,\mathbb{E}_{\lambda\leq\lambda_{\max}} f(\lambda,t)+\frac{1}{2}\aavg{f}~~,
\eeq
where
\beq
\label{eq:2bra}
\aavg{f}=\int_{x_{\min}}^{\lambda_{\max}} \Big[\mathbb{E}_{\lambda\leq x} f(\lambda,t)\Big] \pi(x) dx
\eeq
and $\pi(x)$ stands for the probability distribution of the threshold $x$. Specifically, $\pi(x)=\delta(x-x_{\min})$ in the const-$x$ case and $\pi(x)=(\lambda_{\max}-x_{\min})^{-1}$  for $x\in[x_{\min},\lambda_{\max}]$ in the rand-$x$ case. Note that, because $p(\lambda)\simeq q(\lambda)$ and $f=\lambda$ (resp. $f=0$) for $\lambda<x $ (resp. $\lambda>x$), we have 
\begin{multline}
\mathbb{E}_{\lambda\leq x} f(\lambda,t) \simeq \int_0^{x} \lambda q(\lambda)  d\lambda =\\= \frac{\lambda_{\max}}{(a+2)} \left[ 1 - \left(1-(a+1)\frac{x}{\lambda_{\max}}\right)\cdot\left(1-\frac{x}{\lambda_{\max}}\right)^{a+1}\right]~~.
\end{multline}
Substituting this into (\ref{eq:2bra}) and then in (\ref{Lamd}) one obtains
\begin{widetext}
\beq
\label{eq:Dinf}
\Lambda \simeq \begin{cases}
\displaystyle
\frac{\lambda_{\max}}{(a+2)}\left[ 1 - \frac{(a+2)}{2}\frac{\xm}{\lambda_{\max}}\left( 1 - \frac{\xm}{\lambda_{\max}}\right)^{a+1} - \frac{1}{2}\left( 1 - \frac{\xm}{\lambda_{\max}}\right)^{a+2} \right] & \mbox{(const-$x$ environment)}\\
\displaystyle
\frac{\lambda_{\max}}{(a+2)}\left[ 1 - 
\frac{1}{2}\frac{x_{\min}}{\lambda_{\max}}\left( 1 - \frac{x_{\min}}{\lambda_{\max}}\right)^{a+1} -
\frac{1}{(a+3)}\left( 1 - \frac{x_{\min}}{\lambda_{\max}}\right)^{a+2}
\right] &\mbox{(rand-$x$ environment)} \\
\end{cases}~~.
\eeq
\end{widetext}
These formulas confirm the intuitive picture according to which the more the underlying distribution of phenotypes $q$ concentrates on small values of CRR (i.e. the larger the value of $a$), the slower the population grows at fast phenotypic diffusion. Fig.~\ref{fig:d_a0_20_p03}c,f (horizontal lines at $D\gg 1$) show that the agreement between the long term population growth rate computed numerically and the theoretical estimate given above is excellent in both const-$x$ and rand-$x$ environments.

\subsection{Case of asymmetric switching times}\label{sec:uneven}

We have so far assumed that the characteristic times for switching between selective and non-selective environments are identical. This leaves a single environmental timescale in the problem and simplifies the analysis thanks to the fact that the population spends on average half the time in the selective regime and the other half in the non-selective one. We now want to address the extension of our results to asymmetric switching times.

\begin{figure*}[]
\centering
\includegraphics[ width= 0.95\textwidth]{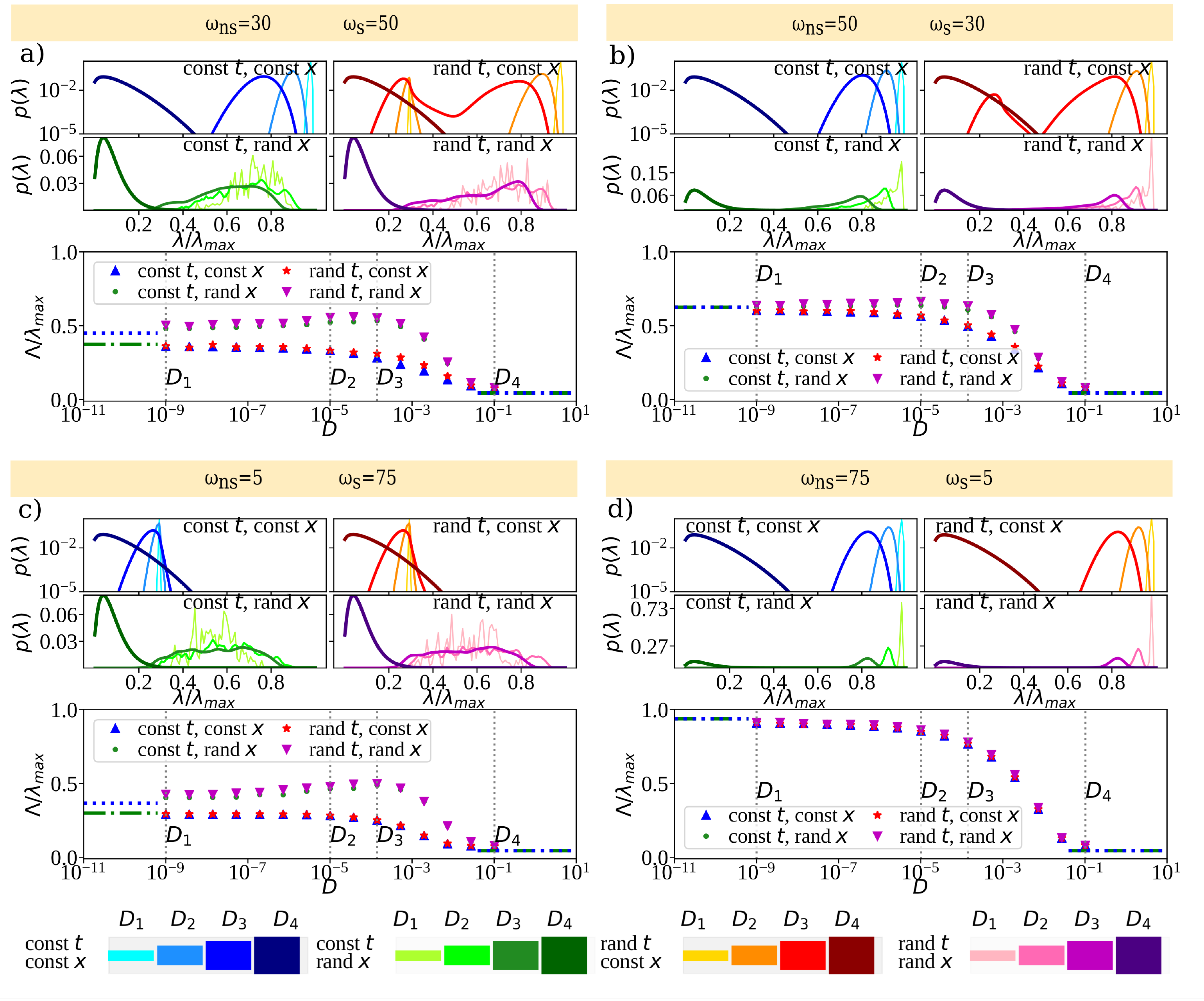}
\caption{\textbf{(a)} Asymptotic, time-averaged phenotypic distributions (top panels) and asymptotic population growth rate $\Lambda$ (in units of $\lambda_{\max}$) as a function of $D$ (bottom panel) obtained as in Fig.~\ref{fig:d_a0_20_p03}d-f but in presence of an asymmetric environment with characteristic switching times  $\wu = 30$ and $\wf= 50$ time units. As in Fig.\,\ref{fig:d_a0_20_p03}, $\xm = 0.3\lambda_{\max}$ and results obtained for different values of $D$ are marked by increasing color shades and line widths. Dotted blue and dot-dashed green horizontal lines at small and large $D$ show the analytical estimates of $\Lambda$ in the exploration and exploitation limits, obtained in the const-$x$  and rand-$x$ regimes, respectively. \textbf{(b to d)} Same as (a) but with different choices of $\wf$ and $\wu$. Displayed curves are averaged over 100 independent realizations of the dynamics.}
\label{fig:asym}  
\end{figure*}

Numerical results (see Fig. \ref{fig:asym}) reproduce the qualitative picture derived in the symmetric case, with some (noteworthy) modifications.
In first place, when the mean time spent in the non-selective environment is larger, the advantage provided by diffusion in complex environments is diminished while the exploitation limit yields higher fitness with respect to the symmetric case. Viceversa, exploration can be tuned to obtain a higher fitness for the population when the mean time spent in the selective environment is larger. The fitness achieved in the exploration limit is however smaller than the symmetric case. Perhaps most interestingly, in two-state environments with random switching times (rand-$t$, const-$x$) the population can still structure in a bimodal fashion, but the weight of the slower part of the distribution (smaller $\lambda$) reflects the (mean) time spent in the selective state (i.e. it increases with $\wu$). This behavior fully corresponds to the classical `bet-hedging'  scenario described e.g. in \cite{gssk}. In other types of environments, though, other population structures are favored.

The key that allows to easily generalize the fast and slow diffusion limits lies in the observation that, instead of spending on average half the time in each environmental state (selective/non-selective), the population now spends a fraction $\pf = \frac{\wf}{\wf+ \wu}$ of time in the non-selective state and a fraction $\pu = 1- \pf$ of time in the selective one. Therefore, the time average of $f$ in the const-$x$ environment (\ref{eq:fc}) now reads
\begin{equation}
\label{eq:fc_asym}
\mathbb{E}_t f(\lambda,t) \simeq 
\begin{cases}
\lambda & \mbox{if}~\lambda<\xm\\
\pf\lambda & \mbox{if}~\lambda\geq \xm~\\
\end{cases}~~.
\end{equation}
The mean IRR displays again a discontinuity at the threshold $\xm$, but now the value of $\lambda$ for which it attains a maximum depends on both $\xm$ and $\pf$. In specific, for $\xm>\pf\lambda_{\max}$ (resp. $\xm < \pf\lambda_{\max}$), the mean IRR has a maximum for $\lambda = \xm$ (resp. $\lambda = \lambda_{\max}$), where $\mathbb{E}_t f(\lambda,t)=\xm$ (resp. $\mathbb{E}_t f(\lambda,t)=\pf \lambda_{\max}$).

Analytical approximation for the population fitness $\Lambda$ that account for asymmetry in the environment can be easily obtained along the lines of Sections \ref{ssec:envlim} and \ref{sssec:highLim}. In particular, in the exploitation limit and with a two-state (const-$x$) environment, expressions (\ref{ventitre}) and (\ref{dimeq}) generalize to
\begin{eqnarray}\label{ventitre_asym}
\Lambda &\simeq& \pf\, \mathbb{E}_{\lambda\leq\lambda_{\max}} f(\lambda,t) + \pu \,\mathbb{E}_{\lambda\leq x_{\min}} f(\lambda,t)\\ & \simeq & \begin{cases}
x_{\min} & \mbox{if } x_{\min} > \pf \lambda_{\max}\\
\pf\lambda_{\max} & \mbox{if } x_{\min} \leq \pf \lambda_{\max}
\end{cases}~~.\label{dimeq_asym}
\end{eqnarray}

Likewise, in the exploitation limit for the rand-$x$ case one finds that Eq.~(\ref{eq:fu}) takes the form
\begin{equation}
\label{eq:fu_asym}
\mathbb{E}_t f(\lambda,t) \simeq 
\begin{cases}
\lambda & \mbox{if}~\lambda<\xm\\
\lambda\left( 1 - \pu\phi\right) & \mbox{if}~\lambda\geq \xm
\end{cases}~~.
\end{equation}
One sees that $\mathbb{E}_t f(\lambda,t)$ now attains a maximum value $\ovl{f}_{\max}$ given by
\beq
\label{eq:max_fu_asym}
\ovl{f}_{\max} = \frac{1}{4 \pu}\frac{\left(\lambda_{\max} - \pf\,x_{\min}\right)^2}{\lambda_{\max} - x_{\min}}~~,
\eeq
at $\lambda = \frac{\lambda_{\max} - \pf\,x_{\min}}{2 \pu}$ if  $x_{\min}<\frac{\lambda_{\max}}{1+\pu}$, while $\ovl{f}_{\max}=x_{\min}$ at $\lambda = x_{\min}$ if $x_{\min}>\frac{\lambda_{\max}}{1+\pu}$. As before, $\Lambda\simeq \ovl{f}_{\max}$.

Finally, in the exploration limit asymmetric environments turn Eq.~(\ref{Lamd}) into
\beq\label{Lamd_asym}
\Lambda \simeq \pf\,\mathbb{E}_{\lambda\leq\lambda_{\max}} f(\lambda,t)+\pu\aavg{f}~~,
\eeq
which allows to generalize Eq.~(\ref{eq:Dinf}) as
\begin{widetext}
\beq
\label{eq:Dinf_asym}
\Lambda \simeq \begin{cases}
\displaystyle
\frac{\lambda_{\max}}{(a+2)}\left[ 1 - \pu (a+2)\frac{\xm}{\lambda_{\max}}\left( 1 - \frac{\xm}{\lambda_{\max}}\right)^{a+1} - \pu\left( 1 - \frac{\xm}{\lambda_{\max}}\right)^{a+2} \right] & \mbox{(const-$x$ environment)}\\
\displaystyle
\frac{\lambda_{\max}}{(a+2)}\left[ 1 - 
\pu\frac{x_{\min}}{\lambda_{\max}}\left( 1 - \frac{x_{\min}}{\lambda_{\max}}\right)^{a+1} -
\frac{2 \pu}{(a+3)}\left( 1 - \frac{x_{\min}}{\lambda_{\max}}\right)^{a+2}
\right] &\mbox{(rand-$x$ environment)} \\
\end{cases}~~.
\eeq
\end{widetext}
Fig.~\ref{fig:asym} (see green and blue horizontal lines) shows that the above expressions for $\Lambda$ provide an excellent agreement with numerical results in both the exploration and exploitation limits.

\section{Discussion}

Empirical data on phenotypic distributions, quantified e.g. from protein expression data, display a rich spectrum of behaviors ranging from unimodal to bimodal depending on the applied stress, organism, etc. (see e.g. \cite{tani} for evidence regarding {\it E.coli}). The question of when one type of distribution is favored therefore appears to be subtle and possibly requires a case by case answer. Our results are in line with previous work in suggesting that the population structure is tightly linked to the specific features of the environment. In particular, when the strength of the coupling between the environment and phenotypes takes on two distinct levels (e.g. high/low, corresponding to the const-$x$ case), bimodal distributions arise but exploration does not yield a fitness advantage to the population. On the other hand, under the more complex scenario in which the coupling strength varies randomly (rand-$x$ case), the exploration-exploitation trade-off leads to a non-zero optimal search rate and unimodal phenotypic distributions are generically preferred. This picture is in complete agreement with the results obtained in \cite{gd}, where the theoretical benefit of a bimodal distribution of stress response proteins was found to be highest in two-state environments, while more variable and structured environments allow for the selection of unimodal distributions. In addition, we have found that adding a small amount of diffusion to a purely exploitative strategy always leads to an increase of fitness in rand-$x$ environments, while it is always detrimental in const-$x$ environments. (More generally, diffusion appears to be broadly beneficial in rand-$x$ environments.) Therefore, both the way a population is distributed across its phenotypic space  and its fitness directly reflect its history in coping with the random environment.

At the quantitative level, the fitness gain given by exploration also appears to be linked to the structure of the underlying phenotypic landscape $q(\lambda)$. In particular, in the more realistic case in which $q(\lambda)$ is strongly heterogeneous, with rare fast phenotypes among a multitude of slow ones \cite{dcd}, a diffusive search dynamics can provide a significant fitness advantage. More generally, it appears to be possible to set the exploration rate within an optimal range for any environment when (i) losses caused by fast diffusion (high $D$) are avoided, while (ii) losses that are to be faced by exploring the phenotypic landscape in two-state (const-$x$) environments are not too large with respect to the $D\to 0$ limit. A rather broad range of values of $D$ fits this criterion, suggesting that, while possibly helpful in certain conditions, a tight regulation of the phenotypic exploration rate may be unnecessary as long as the key assumptions made here hold. 

From a physical viewpoint, our model ultimately relies on Markovianity and detailed balance. These ingredients provide in our view the most elementary way to encode for the effects of fully unbiased random changes in cellular physiology at the level of a complex macroscopic parameter such as the growth rate. However, they are likely to fail in many biologically realistic contexts and moving beyond them would be important. Another limiting modeling choice we made concerns the assumption that faster-growing cells susceptible to environmental shocks do not replicate, as we are implicitly postulating that they survive the shock. While this may be unrealistic in some situations, we note that the introduction of an explicit cellular death rate would effectively re-scale the `replicator' term in (\ref{eq:masterP}). The qualitative scenario we describe should therefore persist. Finally, we focused on a diffusive transition kernel in which only small changes in CRR are allowed. While this is a reasonable choice in biological contexts when significant phenotypic re-arrangements can occur the emergent scenario may be different. For instance, this is likely to be the case when transition rates follow a Gibbs kernel, in which the  $W(\lambda\to\lambda')$ affecting (\ref{eq:master}) is proportional to the density of states with CRR $\lambda'$, i.e.
\beq
\label{eq:wgibbs}
W(\lambda  \rightarrow \lambda') = \frac{q(\lambda')}{\tau}~~.
\eeq
In particular, in this situation diffusion may turn out to be more efficient in improving population fitness than under a diffusive kernel, most notably so in homogeneous landscapes. On the other hand, justifying a kernel like (\ref{eq:wgibbs}) for biological modeling would necessarily require assumptions more extreme, and possibly less realistic, than those made here.

At a more speculative level, this work could shed some light on the origin of phenotypically heterogeneous cell populations such as tumors and may point to educated strategies to control their  diversity. For instance, more heterogeneous populations are more likely to evolve in complex environments, suggesting e.g. that higher intratumoral heterogeneity may be the result of highly variable microenvironments. On the other hand, if the `shocks' are taken to be caused by a therapeutic protocol, our study suggests that subjecting the population to a single repeated dose is effective in quenching its fitness irrespective of the timing of administration.

\appendix

\section{Derivation of Eq.\,(\ref{eq:Pdiff})}\label{app:FP}

We start by noting that
\beq
\label{eq:timeDer}
\dot{n}(\lambda, t) = \dot{p}(\lambda,t)N(t) + p(\lambda,t)\dot{N}(t)~~,
\eeq
where, from Eq.~(\ref{eq:master}), 
\beq
\frac{\dot{N}(t)}{N(t)}=
 \int_0^{\lambda_{\max}} f(\lambda,t)p(\lambda,t) d\lambda = \mathbb{E}_{\lambda\leq\lambda_{\max}} f(\lambda,t)~~.
\eeq
A comparison between Eqs~(\ref{eq:master}) and (\ref{eq:timeDer}) immediately yields the first term on the r.h.s of Eq.~(\ref{eq:Pdiff}). To get the second term, we assume a diffusive transition kernel. Making use of the detailed balance condition (\ref{eq:DetB}) one finds \begin{multline}
\int \Big[W(\lambda' \to \lambda)p(\lambda',t) - W(\lambda \to \lambda')p(\lambda,t)\Big]d\lambda'   \\
 = W( \lambda  \to \lambda + \delta\lambda)q(\lambda)\left[ \frac{p(\lambda + \delta\lambda,t)}{q(\lambda + \delta\lambda)} - \frac{p(\lambda,t)}{q(\lambda)}  \right]  \\  +W( \lambda  - \delta\lambda\to \lambda )q(\lambda -\delta\lambda)\left[ \frac{p(\lambda - \delta\lambda,t)}{q(\lambda - \delta\lambda)} - \frac{p(\lambda,t)}{q(\lambda)}  \right]\\
\simeq \frac{(\delta\lambda)^2}{2\tau} \frac{\partial}{\partial\lambda}\left[q(\lambda)\frac{\partial}{\partial\lambda} \frac{p(\lambda,t)}{q(\lambda)}\right]~~,
\end{multline}
where the last step follows after a second-order expansion in $\delta\lambda$ and we imposed that transitions from $\lambda$ to $\lambda \pm \delta\lambda$ happen with the same  probability (implying that $W(\lambda\to\lambda\pm\delta\lambda)=(2\tau)^{-1}$, see (\ref{eq:tau})). Defining  $D = \frac{(\delta\lambda)^2}{2\tau}$, the second term in Eq.\,(\ref{eq:Pdiff}) is immediately recovered.

\end{document}